# Electrically reduced optical switching threshold of VO2-based THz metasurface

V. E. Kaydashev[1,2], A. S. Slavich[1], I. K. Domaratskiy[1], S. S. Zhukov[1], R. Kirtaev[1], D. A. Mylnikov[1], M. E. Kutepov[2], E. M. Kaidashev[2]

[1]Center for Photonics and 2D Materials, Moscow Institute of Physics and Technology, 141701, Dolgoprudny, Russia
[2]Laboratory of Nanomaterials, Southern Federal University, 344090, Rostov-on-Don, Russia

**Abstract**

The VO2-based THz metasurface optical switching threshold is reduced by more than five orders of magnitude to 0.32-0.4 W/cm2 due to the supply of a subthreshold electric current. Moreover, additional deposition of Au nanoparticles can further reduce the threshold by 30% due to plasmonic effects. The proposed designs and a combined approach to controlling the sub-THz/THz transmission by electric current and light make it possible to overcome the existing bottleneck in design of dynamically programmable THz mirror array based on metasurfaces.

Keywords: $VO_2$, THz metasurface, optical switching threshold, electrical and optical switching, Au nanoparticles, THz mirrors

## Introduction

Next-generation 6G telecommunications systems and new super-resolution THz imaging instruments based on "single pixel camera" paradigm [1] need new tools to in-situ manipulate a wavefront of sub-THz/THz radiation. By using a "static" metasurface made of metal antenna array it is already possible to obtain a great variety of devices for manipulating sub-THz / THz wavefront, i.e. focus radiation [2], obtain beam steering at a given angle [3], alter polarization [4,2], filter wavelength [1] or achieve ultrafast modulation [5]. Due to the use of metal elements the functions of these devices are predetermined by their design. However, the market urgently needs new metasurfaces whose characteristics or even functions are dynamically programmed using electrical or/and optical signals.

$VO_2$-based metasurfaces whose sub-THz transmission/reflection properties are dynamically switched or tuned by isolator-to-metal transition (IMT) do offer great flexibility in programming device characteristics by local heating, electric current or laser light exposure. Some progress in this direction has already been achieved [6,7].

In particular, the C-shaped $VO_2$ slot antenna array switched between "on" and "off" states due to heating of the entire substrate above the transition temperature Tc (~65 °C) and focusing of sub-THz/THz waves or diffraction-free Airy beam formation was obtained in the "on" state [3]. The $VO_2$ film is heated by current and the transmission of $VO_2$ film/Au split-ring metasurface and the position of the resonance change [8]. The THz beam amplitude modulator based on $VO_2$ matasurface was globally electrically switched to become a pair of crossed metallic polarizers [9]. Also, theoretical prediction suggests that an array of $VO_2$-based microstructures connected in parallel can be electrically switched to highly conductive state and THz filtering/polarization rotation can be achieved [10,11]. Several "bit-coded" $VO_2$ metasurfaces capable to dynamically "programming" the phase distribution within the THz wavefront to achieve THz beam steering have also been numerically studied [12,13].

The next step towards designing a dynamically programmable THz metasurface is to achieve an addressing, i.e. to obtain a switching not the entire surface but its individual "pixels" using electrical signals or/and by projecting programmed light pattern. A traditional barely electrical addressing to *N* elements of two dimensional array force one to use many crossed conducting wires which may negatively affect THz properties of a metasurface. An optical switching approach is considered as much more elegant and flexible way to control a device. Unfortunately, as we show in this study for purely optical switching of $VO_2$ "pixels" the incident light intensity should be quite great. Our estimates suggest that purely optical switching of $VO_2$ is achieved at irradiance of visible/near IR light as high as $\sim 7.3\times10^5$ W/cm$^2$ which is hardly compatible with metasurface. Indeed, to optically control a several-mm-scale metasurface one should use either unfocused beam of a several watts cw laser or expensive pulsed lasers. Note, that tight focusing of laser beam to sub-micrometer spot is hardly suitable for simultaneous controlling of many pixels in large metasurface. Therefore, the high optical switching threshold remains a bottleneck for optically addressed $VO_2$ metasurfaces.

Optical addressing performance can be improved by additionally applying electric current to $VO_2$. In particular, by controlling the electric current that is supplied to all $VO_2$ "pixels" connected in parallel we increase their temperature to a value just below the IMT point $T_c$. Such pre-heated "pixels" are easily switched between isolating and metallic state when exposed to continuous laser light, even at low intensities. Is should be emphasized that tight focusing of the beam is no longer needed and a spot of a few millimeters can easily switch the entire metasurface.

In this paper, we study an electrically controlled $VO_2$-based metasurface optical switching threshold reduction to manipulate the wavefront of a transmitted sub-THz/THz beam using low-intensity near IR or UV laser. In addition, we use plasmonic Au nanoparticles to further reduce the switching threshold. The alteration in THz/mid IR/near IR transmission, electric and structural properties caused by direct heating, exposure to laser light and the combined "current-light" protocol are thoroughly investigated. The demonstrated proof-of-concept on optical switching of $VO_2$ metasurface elements opens up a way to dynamically program reflecting metasurfaces in sub-THz range a light pattern in the UV/visible/near IR ranges.

## 2. Experimental details

Epitaxial $VO_2$ films were prepared on a-$Al_2O_3$ (Sample A) and r-$Al_2O_3$ (Sample B, C and D) substrates by pulsed laser deposition method. Briefly, radiation of KrF laser (248 nm, 10 Hz) was focused on a rotating metal vanadium target to obtain a fluence of 2.3 J/cm$^2$. The substrate was positioned at a distance of 5 cm from the target and heated to a temperature of 600°C. The films were deposited at an oxygen pressure of $6\times10^{-2}$ mbar.

First, a buffer layer was deposited for 700 laser shots. Then the deposition was suspended for 1 minute to improve the crystalline quality of the prepared film. Finally, the main film was deposited for another 3300 laser shots. Thereafter, the film was in-situ annealed for 15 minutes at 600°C and cooled to room temperature in an atmosphere of ambient oxygen. Au nanoparticles were additionally deposited on Sample D by pulsed laser deposition at Ar pressure of 0.7 mbar by the method described elsewhere [24,25]. In short, a radiation of KrF laser (248 nm, 10 Hz) was focused on a rotating Au target to produce a fluence of 2 J/cm$^2$. The substrate was positioned at a distance of 3.5 cm from the target. The deposition of nanoparticles was carried out at room temperature for 300 laser shots. The films exhibit excellent structural, electrical and optical switching charactrtistics associated with temperature-induced insulator-to-metal transition that occurs at $\sim$ 55-70 °C for various samples. The changes in the electronic structure of $VO_2$ caused by temperature results in abrupt decrease in the resistance of the film from $\sim$1 MΩ·cm to 100 Ω·cm as shown in Fig.1(a), accompanied by a typical gradual change in the near IR/middle IR as well as sub-THz/THz reflection and transmission properties as shown in Fig.1(b,c).

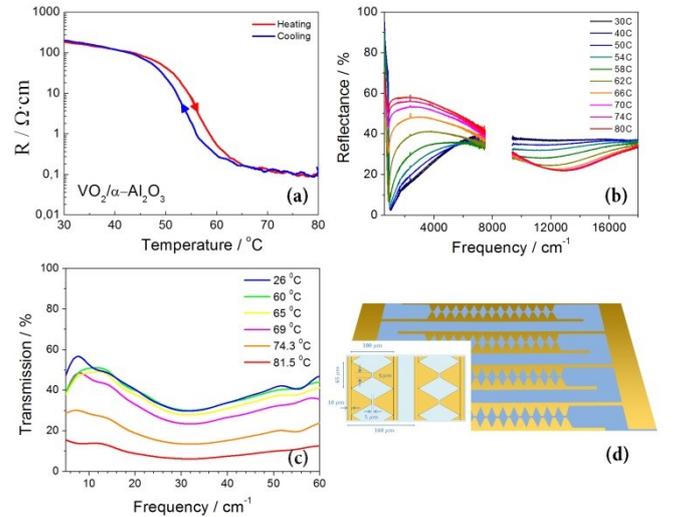

Fig. 1. Temperature induced $VO_2$ film resistance (a), near and middle infrared reflectance (b) and sub-THz/THz transmission (c) alteration. The data presented describe the properties of the $VO_2$ film in Sample A; schematic description of the prepared Au-$VO_2$ metasurface (d).

The sub-THz/THz transmission of as-prepared $VO_2$ film is gradually decreased over a wide frequency range of 0.1-2 THz ($\sim$3-60 cm$^{-1}$) at temperatures above the IMT transition when the film becomes highly conductive. More specifically, the transmission of the film recorded at a frequency of 1 THz is altered from $\sim$33% to 6% as shown in Fig.1(b), which is in qualitative agreement with results presented elsewhere [14]. In addition, the middle IR reflectance of the film is increased as the temperature rises from 30°C to 80°C as shown in Fig. 1(c). In particular, the reflectance measured at 1000 cm$^{-1}$ ($\lambda$=10 $\mu$m) is altered from 2.5% to 56%. The reflection spectra in the near IR range, on contrary, show a decrease from 37% to 22% at $\lambda$=800 nm, which is in a good agreement with the data of other studies [15,16].

The Au(40 nm)/Ti (10nm) bow-tie antenna array shown in Fig. 1(d) was fabricated on the surface of a $VO_2$ film using magnetron sputtering and optical lithography. Two metasurfaces $\sim$2.8 mm in size consisting of 575 (Sample B) 118×78 μm and 774 (Sample C, D) 100×65 μm bow-tie antennas as shown in Fig.2. The separation gap in bow-ties was designed to be 5×5 μm in all samples.

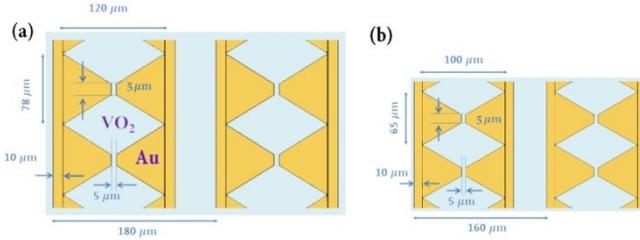

Fig.2. Geometry of fabricated metasurfaces in Samples B (a) and C,D (b)

Note, that antennas in Sample B are 1.2 times larger than those in Sample C. In addition, several series of Au metasurfaces were prepared with scaled sizes of bow-ties on $SiO_2$ (300 nm)/p-Si (12 Ω·cm) substrates and their THz properties were examined.

Sub-THz/THz transmission spectra of metasurfaces with various bow-tie antenna sizes are shown in Fig. 3. The incident THz electric field was chosen to be parallel to the long axis of the bow-ties. The studied metasurfaces reveal the minimum transmission in the range of 0.85-1.36 THz, which is monotonically altered with scaling antenna sizes. All the spectra exhibit significant asymmetry of shape with broad shoulder at higher frequencies. The transmission spectrum of metasurface with 100×65 μm antennas called "M" in Fig.3a has full width at half maximum (FWHM) of ~1 THz. By scaling the bow-ties by 1.2 or 1.4 times to larger

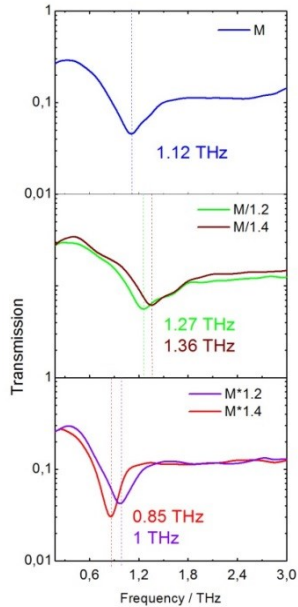

Fig. 3. Sub-THz/THz transmission spectra of Au metasurface on $SiO_2$/Si substrate. Spectrum of Au/$SiO_2$/Si metasurface with geometry similar to the one in Sample C (a), scaled 1.2 or 1.4 times to smaller (b) and larger (c) size. Geometry Au metasurface marked as M×1.2 is similar to one in Sample B.

FWHM to 0.81 THz or 0.56 THz, correspondingly. Likewise, the metasurfaces with bow-ties that are 1.2 or 1.4 times larger show resonant frequency shifted to 1.27 or 1.36 THz and the FWHM increased to 1.21 or 1.24 THz, respectively. Metasurfaces B and C, D are electrically equivalent to an array of 575 and 774 Au-Ti-$VO_2$-Ti-Au structures, which are connected in parallel and located between grating wires. Typical voltage-current characteristics of Au-$VO_2$ metasurface reveal slightly rectifying behavior as shown in Fig.S1 of Supplementary material and, thus, each bow-tie structure is equivalent to couple of diodes connected to each other.

Structural alteration due to the phase transition in the $VO_2$ film induced by heating and tightly focused laser radiation were characterized using Horiba Lab RAMHR Evolution Raman spectrometer. Raman spectra of $VO_2$ films were excited by 532 nm laser light focused by a ×100 objective (NA=0.9) into a 0.4 μm spot. Laser irradiance was controlled in the range $0.74×10^5$ - $14.7×10^5$ W/$cm^2$ using an optical attenuator. Reflection spectra in the middle and near IR were studied using Bruker Vertex 80V FTIR spectrometer equipped with a Hyperion 2000 microscope. Sub-THz/THz transmission spectra of $VO_2$ films, Au bow-tie/$SiO_2$/Si and Au bow-tie/$VO_2$/$Al_2O_3$ metasurfaces were recorded using Tera View TPS Spectra 3000 and Menlo time-domain THz spectrometers with femtosecond laser pumping.

## 3 Results and discussion

### 3.1 Temperature induced structural transition

Raman scattering from $VO_2$ film (Sample A) was recorded at various temperatures ranging from 30 to 80 °C. The laser irradiance was maintained at $0.74×10^5$ W/$cm^2$ to avoid light induced heating. Raman spectra obtained at 25 °C reveal modes at 194, 222, 304, 392, 497, 618 $cm^{-1}$ and at 432, 449, 595 $cm^{-1}$ which are attributed to the $Ag_n$ and $Bg_n$ vibrations of the M1 monoclinic lattice as shown in Fig.4(a) [17-24]. Group theory predicts that monoclinic structure M1 with the space group $C_{2h}^5$=P21/c(14) has 9Ag and 9 Bg Raman active modes. More detailed information on $VO_2$ lattice vibrations assigned in present and previous studies is summarized in Supplementary material Table ST1. It was found that all modes of the M1 phase are significantly suppressed or disappear in the spectra obtained at a temperature of 60 °C and above. Minor traces of another group of modes, namely, at 246, 263, 272, 337 and a shoulder near 640 $cm^{-1}$, which were assigned to the M2 monocline phase. Note that the narrow modes at 124, 378, 437 and 451 $cm^{-1}$ do correspond to lattice vibrations of a single crystal $Al_2O_3$ substrate. At temperatures of 55-70 °C (varies for samples grown on a-$Al_2O_3$ (Sample A) and r-$Al_2O_3$ (Sample B-D) substrates) a phase transition occurs and lattice vibration intensities characteristic of the M1 phase at 199, 222, 246, 263, 272 and prepared VO$_2$ film (Sample A). As shown in Fig. 4(b), the Raman spectra recorded at a laser radiation intensity of more than ~7.3×10$^5$ W/cm$^2$ already exhibit characteristic features associated with the rutile phase, which are similar to those observed in Fig. 4(a) when sample is heated to a temperature above IMT point $T_c$. According to our estimates, the threshold for optical switching of as-prepared VO$_2$ film is in the range of 2.9 – 7.3 ×10$^5$ W/cm2. This high threshold is a bottleneck that limits use of vanadium dioxide for design of metasurfaces which are configured only by optical signals.

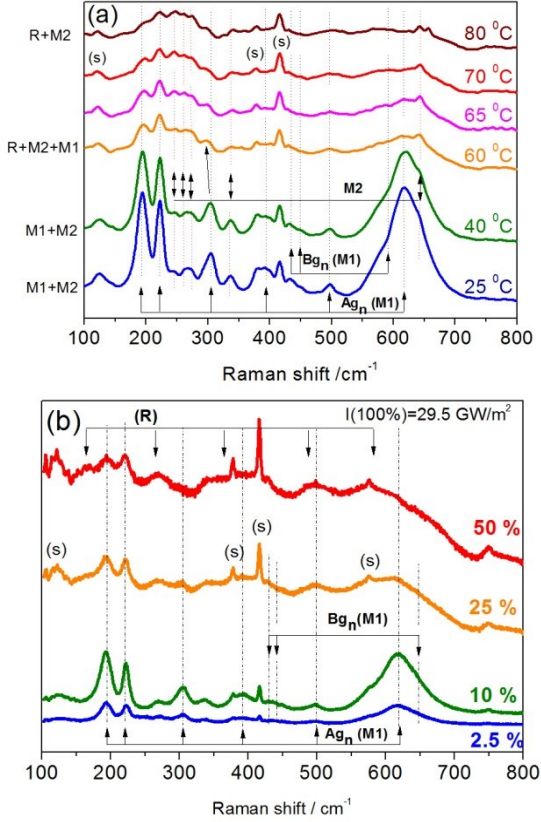

### 3.3 Electrically induced reduction of optical switching threshold of VO$_2$-based metasurface

Alternatively, the phase transition in the VO$_2$ film can be triggered by an applied electric current due to release of Joule heat. Indeed, when a voltage of 3.74 V and 4.28 V is applied to prepared metasurfaces in sample B and sample C, a total current of ~96 mA and 69 mA flows through all its 575 and 774 Au-Ti-VO2-Ti-Au structures, correspondingly. With a slight increase in voltage to 3.78 V and 4.3 V, the insulator-to-metal transition triggers an abrupt increase in current to 174 mA or 224 mA, respectively, as shown in Fig. 5(a). We use this electric current effect to gradually reduce the optical switching threshold of the VO$_2$ metasurface. Indeed, once the applied current is selected to be just below the threshold value, the metasurface "pixels" are easily switched from insulating to metallic state when irradiated with even low-intensity near infrared (808 nm) or ultraviolet (404 nm) laser light as shown in Fig. 5(b,c). More specifically, by applying a constant current of 96 mA to the metasurface in Sample C, we can switch the entire surface to

Fig. 4. Raman spectra of VO$_2$ film measured at different temperatures (a) and irradiances of incident 532 nm laser light (b)

298 cm$^{-1}$ are gradually decrease. The mode at 497 cm$^{-1}$ is broadened and the intense vibration at 618 cm$^{-1}$ as well as the less intense mode at 304 cm$^{-1}$ disappear completely, while several broad bands of the rutile phase (R) appear in the 315-460 cm$^{-1}$, 460-530 cm$^{-1}$, 530-607 cm$^{-1}$ ranges as shown in Fig.4(a).

Indeed, it is generally believed that the high-temperature metallic state of VO$_2$ has a predominantly rutile structure, which is described by the space group $D_{4h}^{14}$=P42/mnm (136) [23]. It is noteworthy that some traces of the monoclinic phase are resistant to heating of the sample and are observed in all spectra recorded both at temperature below 60 °C, when the film is insulating, and at 80 °C, when electrical as well as sub-THz transmission characteristics (see below) clearly indicate metallic properties.

Also, a new low intensity narrow mode at 325 cm$^{-1}$ is observed and assigned to M2 phase [17]. Besides, the temperature increased to 80 °C results in the shift of M2 phase mode at 640 cm$^{-1}$ to 645 cm$^{-1}$, which is similar to behavior of this vibration reported for VO$_2$ nanobeams [17].

### 3.2 Optical switching threshold of VO$_2$

By recording Raman spectra at various irradiances of 532 nm laser, we estimate the optical switching threshold for as-a metallic state by exposing it even with unfocused 4 mm laser beam. Namely, by applying this current and exposing a surface with 30-60 mW laser light, we observe IMT when irradiance is above the threshold value, or not if intensity is insufficient. The optical switching threshold was found to be 0.4 W/cm$^2$ and 0.32 W/cm$^2$ for near IR (808 nm) and UV (404 nm) light, correspondingly. Note that the obtained electrically assisted optical switching threshold is ~10$^5$ times less than purely optical switching threshold found using Raman scattering in Fig. 4(b).

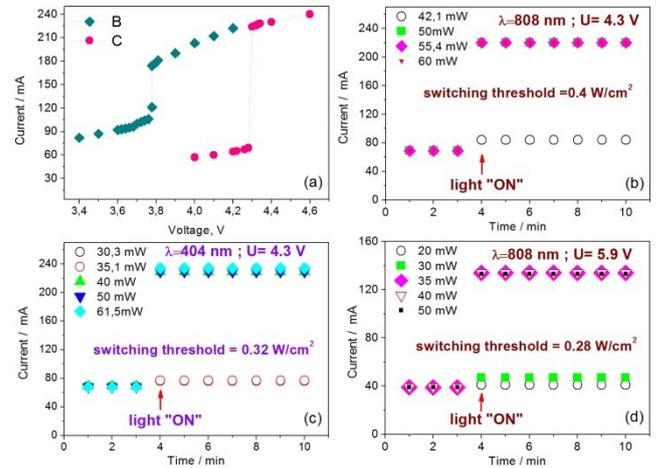

Fig. 5. Electric current induced switching of VO$_2$-based metasurfaces in Samples B and C. Applied voltage of 3.78 V and 4.3 V trigger an insulator-to-metal transition with a drop of resistance (a); switching of metasurface in Sample C with bare VO$_2$ film triggered by an unfocused 4 mm laser beam at 808 nm (b) and UV 404 nm (c) at constant current of 96 mA; switching of metasurface in Sample D with VO$_2$ film and Au nanoparticles triggered by unfocused light at 808 nm when constant current of 42 mA is applied (d).

### 3.4 Reduction of VO$_2$ optical switching threshold induced by plasmonic nanoparticles

To further boost a performance of photoinduced switching, the monolayer of Au nanoparticles was deposited on VO$_2$ film in Sample D with bow-ties similar those in Sample C as shown in Fig. 6. Similar to previous experiment, the metasurface in Sample D was heated with an electric current of 42 mA to approach to IMT. The optical switching threshold is found to be 0.28 W/cm$^2$ (35 mW in 4 mm spot) as shown in Fig. 5d, which is 30% lower compared to 0.4 W/cm$^2$ for Sample C with bare VO$_2$ film. Note, that the slightly higher voltage applied to Sample D is caused by the higher resistance of Au-Ti-VO$_2$ contacts due to technological issues related to contacts quality and should not be considered here. Au nanoparticles ~10 nm in size are similar to those studied elsewere [24-25].

A monolayer of particles has a broad plasmonic absorption band centered at ~750 nm [25]. When illuminated by laser light with wavelength of 808 nm, the particles (i) themselves heat up due to strong plasmon absorption, and also (ii) cause additional optical absorption in the VO$_2$ film. These actions result in a greater photothermal effect in Au-VO$_2$ system, which manifests itself as a lower light-induced switching threshold.

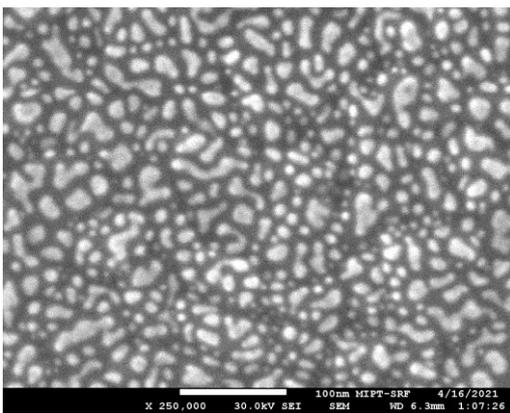

Fig. 6. Morphology of VO$_2$-Au metasurface in Sample D with Au nanoparticles

### 3.5 Optical and electrical switching of THz metasurface

Finally, three different methods of switching the transmission properties of Au-VO$_2$ sub-THz/THz metasurfaces shown in Fig.1d were examined, namely induced by temperature change, application of electric current and a combined method with preheating the surface with current and triggering a phase transition with low intensity laser radiation. A typical spectrum of the VO$_2$-Au structure shown in Fig.7 in the first approximation can be considered as a superposition of the spectra of a VO$_2$ film and a metallic bow-tie array similar to those for Au/SiO2 structure shown in Fig.3. At room temperature, the transmission of the metasurface is monotonically decreased from 47 to 30% in spectral range of 0.1-0.6 THz and reveal a local minimum of 1% at 1.2 THz as shown in Fig.7a. Note that the sub-THz/THz spectra show no noticeable change with temperature changes in the range of 24-66°C (only one spectrum is shown here). Once the temperature is increased to 75°C or higher value, the transmission drops abruptly to 1-4% at all frequencies in the 0.1-2 THz range due to the large reflection of the conductive VO$_2$ film.

A gradual change in THz transmission is also achieved by applying an electric current, as shown in Fig. 7(b), which is similar to temperature induced alteration. Indeed, the THz

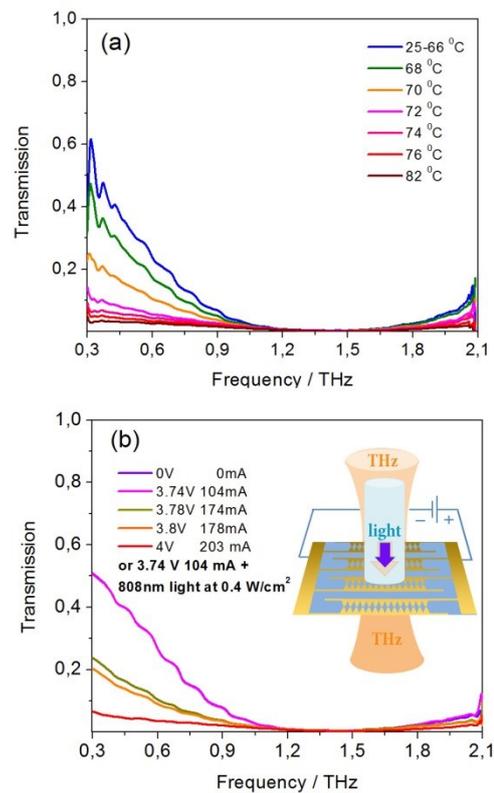

Fig. 7. Alteration of metasurface sub-THz/THz transmission induced by temperature (a), by electric current and by unfocused 808 nm laser light at 0.4 W/cm$^2$ while an optical switching threshold is reduced to 0.4 W/cm$^2$ by additional 104 mA current. Proof-of-concept experiment is shown in the inset. (b)

reflection alteration mimics the change in the IMT related resistance drop shown in Fig. 5(a). In particular, when the voltage is less than 3.78 V (Sample B) and the current does not exceed 104 mA, the THz transmission is not effected. At electric currents above the threshold value of 104 mA, the metasurface sharply becomes highly reflective with respect to THz radiation with a decrease in transmission to 1-4%.

Proof-of-concept experiment of this paper, which illustrates optical switching of a metasurface by low-intensity light due to an additionally applied electric current, is also shown in Fig.7(b). Similar to experiment shown in Fig.5 (b,c), a subthreshold voltage of 3.74 V was applied to Sample B to adjust its temperature just below the IMT point. In such condition the THz transmission of the metasurface is easily switched even by unfocused laser light at 808 nm with intensity of only 0.4 W/cm$^2$ as shown in Fig.7 (b). Note, that transmission curve coincides with one in case of barely electrical switching by current of 104 mA applied to metasurface. The proposed type of metasurface is fully compatible with the concept of light-controlled addressing of the individual VO$_2$ "pixels" in the THz metasurface and will be useful for local programming of the front of transited/reflected THz wave. Projecting of a "light pattern" generated by computer on the studied VO$_2$-Au metasurface is a subject of our further research.

*4. Conclusion*

In summary, the threshold for optical switching of the sub-THz/THz metasurface based on VO$_2$-Au is reduced by ~10$^5$ times due to supply of an electric current, which determines the operating point immediately before the transition of VO$_2$ from insulator to the metal. A further decrease in the optical switching threshold by ~30% is induced by a monolayer of Au nanoparticles. An experiment confirming the concept is demonstrated to illustrate the switching of the broadband transmission of the sub-THz metasurface by an unfocused UV or NIR laser beam of 4 mm with an intensity of only 0.32-0.4 W/cm$^2$. The proposed electrically assisted and optically triggered switching of "THz pixels" opens up a bypass route to overcome a bottleneck in the development of a dynamically programmed THz metasurface.

**Acknowledgement**

The fabrication of metasurfaces and study of metasurfaces` THz properties was supported by Russian Science Foundation, grant № 22-29-01037 at SFU. Study of IR/middle IR reflection and Raman scattering of VO$_2$ films was supported by Russian Science Foundation, grant № 21-79-00209 at MIPT. The deposition and optimization of VO$_2$ films was supported by Southern Federal University project No. 07/2020-06-MM. The metasurfaces were fabricated using equipment of MIPT Shared Facilities Center with financial support from the Ministry of Education and Science of the Russian Federation (Grant No. RFMEFI59417X0014).